\documentclass[a4paper,11pt]{article}
\pdfoutput=1 

\usepackage{jcappub} 
\usepackage[T1]{fontenc} 

\usepackage{natbib}
\usepackage{cancel}
\usepackage{enumitem}    
\usepackage{amsthm}         
\usepackage{amsmath} 	   
\usepackage{amssymb}       
\usepackage{amsfonts}
\usepackage{graphicx} 	    
\usepackage{verbatim}
\usepackage{epsfig,multicol,bbm}
\usepackage{color}
\usepackage{colortbl}
\usepackage{float}
\usepackage{multirow}
\usepackage{array}

\usepackage{subfig}

\usepackage{xcolor}
\usepackage{hyperref}

\allowdisplaybreaks

\title{Sourcing curvature modes with entropy perturbations in non-singular bouncing cosmologies}
\author[a,b,1]{Anna Ijjas,}
\note{Corresponding Author}
\author[c]{Roman Kolevatov}

\affiliation[a]{Max Planck Institute for Gravitational Physics (Albert Einstein Institute), Hannover, D-30167, Germany}
\affiliation[b]{Leibniz Universit\"at Hannover, D-30167 Hannover, Germany}
\affiliation[c]{Department of Physics, Princeton University, Princeton, NJ 08544, USA}

\emailAdd{anna.ijjas@aei.mpg.de}

\abstract{The observed temperature fluctuations in the cosmic microwave background can be traced back to primordial curvature modes that are sourced by adiabatic and/or entropic matter perturbations. In this paper, we explore the entropic mechanism in the context of non-singular bouncing cosmologies. We show that curvature modes are naturally generated during `graceful exit,' {\it i.e.}, when the smoothing slow contraction phase ends and the universe enters the bounce stage. Here, the key role is played by the kinetic energy components that come to dominate the energy density and drive the evolution towards the cosmological bounce.}


\begin{document}
\maketitle 
\flushbottom

\section{Introduction}
Curvature perturbations of the metric created in the very early universe induce the primordial density inhomogeneities that are responsible for the observed temperature fluctuations in the cosmic microwave background (CMB) and the seeds for forming stars and galaxies. The most recent Planck results \cite{Aghanim:2018eyx,Akrami:2018odb,Akrami:2019izv}, together with earlier observations from the Wilkinson Microwave Anisotropy Probe (WMAP) \cite{Komatsu:2008hk} and the Atacama Cosmology Telescope (ACT) \cite{Sievers:2013ica}, show that the spectrum of primordial density fluctuations is adiabatic, nearly scale-invariant, and Gaussian.

While it is an unsettled question {\it when} in cosmic history the curvature perturbations were generated, there is strong observational evidence that the curvature perturbations are sourced by {\it quantum} fluctuations of one or more scalar matter fields that occur on wavelengths smaller than the Hubble radius \cite{Bardeen:1983qw,Sasaki:1983kd,Mukhanov:1988jd}. By one means or another, the wavelengths end up on scales much larger than the Hubble radius, evolve into {\it classical} curvature perturbations, by some time prior to primordial nucleosynthesis, and set the conditions required to explain galaxy formation and CMB temperature fluctuations in the later stages of cosmic evolution. 

Quantum fluctuations of the scalar matter field come in two types: perturbations of the total energy density $\delta \varrho$ on hypersurfaces of constant mean curvature are called `adiabatic;'  and perturbations of the total pressure on hypersurfaces of constant energy density ($\delta\varrho\equiv0$) are called `entropic.' Which one of the two types yields the dominant contribution to the curvature modes depends on the background mechanism that drives the underlying cosmological evolution. For example, during inflation, both adiabatic and entropic modes are excited by the rapid background expansion. Accordingly, in some proposed inflationary scenarios the curvature modes are sourced solely by the adiabatic mode ({\it e.g.}, single-field slow-roll models \cite{Guth:1980zm,Linde:1981mu,Albrecht:1982wi}) and in other scenarios  solely by the entropic mode ({\it e.g.}, curvaton models \cite{Lyth:2001nq}). In yet other inflationary constructions, both adiabatic and entropic modes contribute to the curvature fluctuations \cite{La:1989za}. 

Notably, in bouncing and cyclic cosmological theories where the universe is flattened and smoothed by undergoing a phase of slow contraction that is connected to the current expanding epoch through a cosmological bounce \cite{Khoury:2001wf,Ijjas:2018qbo,Ijjas:2019pyf}, there are not all these options.  In a slow contraction phase,
the amplitudes of adiabatic fluctuations decay{}. On the other hand,
in non-singular bouncing cosmology, for example, in which the background solution is a dynamical attractor \cite{Levy:2015awa},  the amplitudes of entropic modes grow.  Consequently, the entropic modes are inevitably the dominant source of the curvature modes observed in the CMB in these theories. 

One mechanism for generating entropy modes that has been discussed in the literature for either inflationary expansion or slow contraction smoothing phases entails a linear combination of two scalar fields with canonical kinetic energy density rolling down a trajectory on a two-dimensional scalar potential.  In this case, the entropic fluctuations originate from quantum fluctuations orthogonal to the trajectory.   After the smoothing phase completes, the entropy modes can source curvature perturbations if the two-dimensional scalar potential is shaped just as to cause the trajectory to change sharply just as the smoothing phase ends \cite{Khoury:2001wf,Finelli:2002we,Lehners:2007ac,Buchbinder:2007ad,Koyama:2007ag}.  

In the case of classical (non-singular) bouncing and cyclic cosmology, though, tuning a potential in this way appears {\it ad hoc} because here, as the slow contraction phase ends, the kinetic energy of the scalar matter fields naturally comes to dominate over the potential energy as the scale factor continues to decrease.  For these cosmologies, a kinetic energy density driven mechanism for using the entropic fluctuations to source curvature fluctuations would be  the natural approach, in principle. 

In this paper, we show that just such a `kinetic entropic mechanism' is possible and that it can be used to generate a nearly scale-invariant spectrum of curvature perturbations on super-Hubble scales. Here, a key role is played by the same higher-order kinetic energy terms that initiate the classical bounce stage.

The paper is organized as follows. In Sec.~\ref{sec:hydro}, we briefly review and expand on the classic hydrodynamic approach to perturbation theory to derive {\it generic} conditions for sourcing super-Hubble curvature modes from the entropic matter perturbations.  We then obtain the key equation for the time evolution of curvature perturbations that include sourcing by entropy perturbations and discuss a wide range of possibilities for generating a nearly scale-invariant spectrum of curvature perturbations on super-Hubble scales. In Sec.~\ref{sec:micro}, we present a microphysical realization that involves two kinetically coupled scalar fields with a scalar field potential. Most importantly, we derive a master equation for the time variation of the curvature perturbation in terms of the relative field fluctuations. From the latter equation, various microphysical possibilities for generating curvature perturbations on large scales are presented in Sec.~\ref{sec:special}.  Finally, we consider a case in which the generation occurs without any scalar field potential at all, as is natural near a bounce -- that is, a purely kinetic mechanism. Notably, in this case, the generation process depends not only on the relative field perturbation but also on its time derivative. In Sec.~\ref{sec:concl}, we summarize our findings and sketch out possibilities for further studies.

\section{Hydrodynamic analysis}
\label{sec:hydro}

According to our current understanding, the generation of primordial curvature perturbations is closely tied with the classical mechanism that smooths and flattens the cosmological background and drives it to a homogeneous and isotropic Friedmann-Robertson-Walker (FRW) geometry.  
Namely, the curvature fluctuations as observed in the CMB originate from quantum fluctuations  of the stress-energy that occur during the last 60 $e$-folds of smoothing. 

In this section, we shall employ a (macroscopic) hydrodynamic {\it ansatz} \cite{Mukhanov:1990me,Wang:1997cw}  to characterize the background evolution as well as the corresponding adiabatic and entropic fluctuations generated by the quantum fluctuations. The main result of the section is a master equation relating the co-moving curvature perturbation $\mathcal{R}$ to its possible sources, the adiabatic and entropy perturbations.
 
The hydrodynamic approach has the advantage of being both minimalist and generic at the same time.  It is minimalist in that it suffices to reduces the description of the stress-energy to just four relativistic fluid parameters: the equation of state, the sound speed of the adiabatic modes, the non-adiabatic pressure and the anisotropic  stress.  From this, the hydrodynamic ansatz can be used to identify generic predictions for different classes of stress-energy without reference to specific microphysical realizations.

\subsection{Background evolution}

Slow contraction is a `super-smoother,' meaning that it  rapidly homogenizes, isotropizes, and flattens the universe according to the classical equations of motion and also when quantum effects are included,  
 even when starting from initial conditions that lie far outside the perturbative regime of FRW spacetimes \cite{Cook:2020oaj,Ijjas:2020dws}. At the same time, because the Hubble radius shrinks much faster than the scale factor during slow contraction, the wavelengths of quantum fluctuations of the stress-energy generated on scales smaller than the Hubble radius evolve to become much larger than the Hubble radius and classicalize  by the end of the smoothing phase, providing a natural source of primordial curvature perturbations.   

The simple underlying idea is to introduce a form of stress-energy that rapidly becomes the dominant source of energy density during a contracting phase and slows the expansion of the universe so as to drive the geometry towards a FRW universe.
Due to the symmetries of the FRW geometry, this new component acts on the background as a perfect fluid, 
\begin{equation}
T_{\mu\nu} = \big( \varrho + p \big)u_{\mu}u_{\nu} - p g_{\mu\nu},
\end{equation}
and can therefore be characterized through its energy density $\varrho$, pressure $p$, and co-moving velocity $u_{\mu} = (-1, 0, 0,0)$. Note that, as with any perfect fluid, the background pressure $p$ is a function of the energy density 
\begin{equation}
p \equiv  p(\varrho) \equiv \Big( {\textstyle \frac23} \varepsilon - 1 \Big)\varrho.   
\end{equation}
The fluid is fully specified by the dimensionless quantity $\varepsilon$, known as the equation of state, that relates the pressure $p$ with its energy density $\varrho$.   For simplicity, we will treat $\varepsilon$ as approximately constant during the smoothing phase.

The dynamics of the smoothing component follows the continuity equation  
\begin{equation}
\label{eq:hydro_continuity}
\dot{\varrho} + 2\varepsilon H \varrho = 0,
\end{equation}
where $H(t) = \dot{a}/a$ is the Hubble parameter;  $a(t)$ is the FRW scale factor; and dot denotes differentiation with respect to the proper FRW time coordinate $t$. (Here and throughout, we use reduced Planck units.)  We choose the time coordinate such that, during the smoothing phase, $t$ runs from some initial value $t_0 \ll 0$ to some negative $t_{\rm end}$ that is much closer to zero.  We take $a(t_0)=1$. The bounce phase that takes the universe from slow contraction to expansion begins at $t=t_{\rm end}$.   

For constant $\varepsilon$, Eq.~\eqref{eq:hydro_continuity} implies
\begin{equation}
\varrho = \frac{\varrho^0}{a^{2 \varepsilon}}\,,
\end{equation}
where $\varrho(t_0) = \varrho^0$ and the FRW scale factor has been  normalized such that $a(t_0)=1$. 
Note that, according to the continuity equation~\eqref{eq:hydro_continuity},
 the smoothing component is  anti-damped because $H<0$; {\it i.e.}, its energy density is being continuously enhanced by gravity.

To homogenize, isotropize, and flatten the cosmological background, the  energy density of the smoothing component $\varrho$ must come to dominate all other contributions to the  energy density, the anisotropy and the spatial curvature on the right hand side of the generalized Friedmann constraint,
\begin{equation}
\label{FRW-const}
H^2 = \frac13\left( \frac{\varrho^0_m}{a^3} + \frac{\varrho^0_r}{a^4} + \frac{\varrho^0}{a^{2 \varepsilon}}  \right) + \frac{\gamma_s^2}{a^6} - \frac{k}{a^2},
\end{equation}
where  $\varrho_m^0$ and $\varrho_r^0$ are the  (pressureless) matter and radiation densities at $t_0$; and the last two terms characterize the anisotropy and spatial curvature contributions, respectively. It is immediately apparent from this constraint that a necessary condition for the smoothing component to grow to become the dominant contribution to $H^2$ during contraction is that $\varepsilon>3$.   The constraint then reduces to 
\begin{equation}
 3 H^2 \approx \frac{\varrho^0}{a^{2 \varepsilon}}, 
\end{equation} 
driving the universe towards 
 the FRW scaling attractor solution,
\begin{equation}
\label{scaling-sol}
a(t) = (-t)^{1/\varepsilon},\quad H(t) = \frac{1}{\varepsilon\, t}.
\end{equation}
Note that $a(t)$ shrinks more slowly than the Hubble radius, $|H|^{-1} = \varepsilon t$ during the contracting phase.  In fact, the  
 larger is $\varepsilon$, the less the scale factor $a(t)$ (or any physical distance) shrinks for the same change in Hubble radius.  For example, for $\varepsilon =60$, during a period of slow contraction in which the Hubble radius shrinks by $60$ $e$-folds, the scale factor changes by only a factor of two.  

The radically different rates at which the  scale factor and the Hubble radius shrink is  the key to smoothing and flattening the background.  In the example above, for instance, the measure of the cosmic spatial curvature,  $\Omega_k \equiv k/(a^2 H^2) \propto |H|^{-2}/a^2$, shrinks by over hundreds $e$-folds each time $a(t)$ shrinks by a factor of $e$.  

The difference in rates is also critical for obtaining perturbations on super-Hubble scales. 
The {\it wavelength} of any quantum fluctuation that is generated on sub-Hubble scales shrinks at the same slow rate as the scale factor $a$. Because the Hubble radius shrinks at a much faster rate, any sub-Hubble mode quickly ends up on wavelengths much larger than the Hubble radius by $t_{\rm end}$.

The smoothing phase ends, when $\varepsilon$ decreases and falls to (or below) three. This is the stage of {\it graceful exit}, leading up to the bounce stage when the universe transits to the current expanding phase.

\subsection{Scalar perturbations}

Next, we characterize how the {\it amplitude} of adiabatic and entropy fluctuations in the smoothing matter component evolve during slow contraction and relate them to the curvature fluctuations of the metric.

The amplitude of scalar perturbations in the metric and the smoothing stress-energy component evolve according to the linearized Einstein equations and the linearized equations describing the conservation of stress-energy. To represent the scalar metric perturbations in a coordinate independent way, we use the gauge-invariant Bardeen variables \cite{Bardeen:1980kt},
\begin{subequations}
\begin{alignat}{1}
\Psi & \equiv \psi + H \sigma_{\rm sh}
,\\
\Phi & \equiv \alpha - \dot{\sigma}_{\rm sh}
,
\end{alignat}
\end{subequations}
where  $\alpha$ and $\beta$ are the lapse and shift perturbation, respectively; $\delta g_{ij} = -2 a^2(t) (\psi\delta_{ij} + \partial_i\partial_j \eta)$  is the scalar part of the perturbed spatial metric; and 
$\sigma_{\rm sh} \equiv a^2(t) \dot{\eta} - a(t)\beta$ is the scalar part of the linearized shear component.
Similarly, we write the components of the linearized smoothing stress-energy -- the energy density, pressure, and co-moving velocity -- in a coordinate independent way:
\begin{subequations}
\begin{alignat}{1}
\delta \varrho_{\rm N} & \equiv \delta\varrho -  \dot{\varrho}\,\sigma_{\rm sh}
,\\
\delta u_{\rm N} & \equiv \delta u + \sigma_{\rm sh}
,\\
\delta p_{\rm N} & \equiv \delta p -  \dot{p}\,\sigma_{\rm sh}
.
\end{alignat}
\end{subequations}
(The subscript $N$ refers to Newtonian or zero-shear gauge; note that, in this gauge, the linearized scalar field matter variables are automatically gauge-invariant.)
 
 Furthermore, we decompose the linearized pressure component into a part that is proportional to the linearized energy density $\delta\varrho_{\rm N}$ and a residual contribution:
 \begin{equation}
\label{eq:hydro_non-adiabatic_press}
\delta p_{\rm N} \equiv  \frac{\dot{p}}{\dot{\varrho}}  \delta\varrho_{\rm N}  +  \delta p_{\rm nad}\,,
\end{equation}
where the gauge-invariant residual $\delta p_{\rm nad}$ is called the `non-adiabatic pressure' perturbation \cite{Bardeen:1980kt,Wands:2000dp}.
Note that the proportionality factor, $\dot{p}/\dot{\varrho}$, is a formal quantity. In general and especially in the case of scalar fields, it is not equivalent to the propagation (or sound) speed of the perturbed matter field variables, as we will discuss below. The non-adiabatic pressure perturbation $\delta p_{\rm nad}$ is an input parameter that, together with the dynamical variables $\delta \varrho_{\rm N}$ and $\delta u_{\rm N}$, fully specify the linearized stress-energy. Note that, {\it prima facie}, there is no reason why $\delta p_{\rm nad}$ should be zero.

With these variables, we can write the scalar part of the linearized Einstein equations for each Fourier mode with wavenumber $k$ in a coordinate independent way:
\begin{subequations}
\begin{alignat}{1}
\label{eq:hydro_Einst_00}
& 3H\Big(\dot{\Psi} + H\Phi \Big) + \frac{k^2}{a^2}\Psi 
 = - \frac12 \delta \varrho_{\rm N}
,\\
\label{eq:hydro_Einst_0i}
&\dot{\Psi} + H\Phi 
  = -\frac13\varepsilon\varrho  \delta u_{\rm N}
,\\
\label{eq:hydro_Einst_ij}
& \Psi - \Phi  = a^2 \pi^S
,\\
\label{eq:hydro_Einst_ii}
& \ddot{\Psi} + 3H \dot{\Psi} +H \dot{\Phi} +  \left(2\dot{H}+3H^2\right) \Phi
 = \frac{1}{2}\Big(  \delta p_{\rm N} - k^2 \pi^S \Big).
\end{alignat}
\end{subequations}
Here Eqs.~(\ref{eq:hydro_Einst_00}-\ref{eq:hydro_Einst_0i}) are the Hamiltonian and momentum constraints; Eq.~\eqref{eq:hydro_Einst_ij} is the anisotropy equation; and Eq.~\eqref{eq:hydro_Einst_ii} is the linearized pressure equation. Note that, for completeness, we included a non-zero anisotropic stress component $\pi^S$. In many sources of stress-energy, such as a canonical scalar field, $\pi^S\equiv0$, in which case $\Phi =\Psi$.

The linearized conservation equations for the energy density and momentum of the perturbed hydrodynamic matter take the following form:
\begin{subequations}
\begin{alignat}{1}
\label{eq:perturbs_conservation}
& \delta \dot{\varrho}_{\rm N} + 3 H \Big(\delta \varrho_{\rm N}+ \delta p_{\rm N} \Big)  + 6\dot{H}  \dot{\Psi} 
=  -  \frac{k^2}{a^2}\left( 2 \dot{H} \delta u_{\rm N}  - a^2H  \pi^S \right)
,\\
\label{eq:perturbs_conservation-M}
& -2\dot{H} \left(
\delta \dot{u}_{\rm N} - 3 H \frac{\dot{p}}{\dot{\varrho}} \delta u_{\rm N}   + \Phi
\right)  = -  \delta p_{\rm N}   + k^2  \pi^S,
\end{alignat}
\end{subequations}

An observationally relevant quantity is the curvature perturbation on co-moving hypersurfaces $\mathcal{R}$, a quantity that is directly related to the observed fluctuations in the CMB and the primordial density fluctuations that seeded galaxy formation \cite{Bardeen:1983qw}:
\begin{equation}
\label{eq:hydro_R}
\mathcal{R} \equiv \psi - H\delta u.
\end{equation}
Note that, by construction, ${\cal R}$ is gauge invariant.
The linearized Einstein-scalar field equations~(\ref{eq:hydro_Einst_00}-\ref{eq:hydro_Einst_ii}) can be combined to yield a first-order expression for the time evolution of ${\cal R}$:
\begin{equation}
\label{R-dot-eq-full}
- \frac{\dot{H}}{H} \dot{\cal R}  = \frac12 \delta p_{\rm nad} - \frac{k^2}{a^2}\left( \frac{\dot{p}}{\dot{\varrho}}  \Psi + \frac12\big(\Psi - \Phi\big) \right) .
\end{equation}
Note that $ \dot{\cal R}$ is not gauge-invariant (even though $\cal R$ is), because it involves a time-derivative that depends on the time-slicing.  A consequence is that a description of the evolution of $\cal R$ on super-Hubble scales, an unobservable quantity, is different for different gauge choices. 

In particular, in the absence of non-adiabatic pressure ($\delta p_{\rm nad} \equiv 0$), Eq.~\eqref{R-dot-eq-full} implies that, on super-Hubble scales ($aH\gg k$), the amplitudes of the co-moving curvature modes are constant, as was first pointed out by Bardeen, Steinhardt, and Turner in Ref.~\cite{Bardeen:1983qw}. A corollary is that the amplitudes exiting the Hubble radius and re-entering are the same.  As the amplitudes on sub-Hubble scales, including at exit and re-entry, are observable quantities, the corollary is a coordinate independent statement.  However, the statement that the amplitudes were constant when the modes were on super-Hubble scales is gauge-dependent.   
In fact, in certain often-used gauges ({\it e.g.}, in unitary or flat gauge where $\cal R$ obeys a second-order equation), ${\cal R}$ does not remain constant on super-Hubble scales. In these gauges, recovering the fact that ${\cal R}$ is the same at exit and re-entry becomes a rather cumbersome exercise. For details, see Ref.~\cite{Bardeen:1983qw}.

If the non-adiabatic pressure contribution is non-zero ($\delta p_{\rm nad} \neq 0$), co-moving curvature modes are generated by $\delta p_{\rm nad}$ on super-Hubble scales. 
The intriguing fact that non-adiabatic pressure fluctuations act as a source of co-moving curvature modes on super-Hubble scales, which we will exploit below in Sec.~\ref{sec:micro}, can be viewed as a feature of the linearized scalar metric field evolution equation~\eqref{eq:hydro_Einst_ii} when combined with the three constraint equations~(\ref{eq:hydro_Einst_00}-\ref{eq:hydro_Einst_ij}). Alternatively yet equivalently, it can be viewed as a consequence of momentum conservation for hydrodynamic matter within Einstein gravity,  in the sense that Eq.~\eqref{R-dot-eq-full} can be derived by combining the momentum conservation equation~\eqref{eq:perturbs_conservation-M} with the same three constraints of Einstein relativity.

We note that in Ref.~\cite{Wands:2000dp} it was suggested to view the non-adiabatic sourcing mechanism as a direct consequence of the local conservation of stress-energy ($n^\nu\nabla_\mu T^\mu{}_\nu = 0$, where $n^\mu$ is a unit time-like vector) for any relativistic theory of gravity.
Indeed, it is possible to recast Eq.~\eqref{eq:perturbs_conservation} and obtain a conservation equation
\begin{equation}
\label{eq:hydro_zeta-dot}
\dot{\zeta} =-H \,\frac{\delta p_{\rm nad}}{\varrho + p} + \frac13 \,\frac{k^2}{a^2} \left( \delta u_{\rm N} -  \frac{\varrho}{\dot{\varrho}} a^2\pi^S
\right).
\end{equation}
for the curvature perturbation on uniform-density hyper-surfaces
\begin{equation}
\label{eq:hydro_zeta}
\zeta \equiv -\psi - H\frac{\delta\varrho}{\dot{\varrho}},
\end{equation}
only by assuming Lorentz invariance and without invoking the Einstein field equations~(\ref{eq:hydro_Einst_00}-\ref{eq:hydro_Einst_ij}) or any particular theory of gravitation.
However,  the energy density and velocity perturbations, $\delta\varrho_N, \delta u_N$, and hence the curvature perturbations $\zeta$ and $\mathcal{R}$ are  generally distinct and can only be related by invoking the gravitational field equations. In the case of Einstein general relativity, $\zeta$ and $\mathcal{R}$ can be shown to approach one another in the long wavelength limit ($k/aH \ll1$). That is, by using the linearized Hamiltonian and momentum constraints (Eqs.~\ref{eq:hydro_Einst_00}-\ref{eq:hydro_Einst_0i}), we obtain a relativistic generalization of the Poisson equation, 
\begin{equation}
\label{eq:hydro_energy_momentum_constr}
-\dfrac{k^2}{a^2}\Psi = \frac12 \Big(\delta\varrho_{\rm N} +\dot{\varrho} \,\delta u_{\rm N} \Big).
\end{equation}
Combining Eq.~\eqref{eq:hydro_energy_momentum_constr} with the definitions of $\zeta$ and ${\cal R}$ yields the relation 
\begin{equation}
\label{eq:hydro_R_zeta_relation}
\mathcal{R} = -\zeta + \dfrac{1}{3\dot{H}}\dfrac{k^2}{a^2}\Psi,
\end{equation}
meaning that the co-moving curvature perturbations and the curvature perturbations on uniform density hypersurfaces coincide on super-Hubble scales.
We stress, though, that the same is not generally true for modified gravity theories, such as Horndeski gravity \cite{Ijjas:2017pei}.

The non-adiabatic pressure component induces an entropy perturbation $\mathcal{S}$,
\begin{equation}
\label{def-S-hydro}
\mathcal{S} = H\dfrac{\delta p_{\rm nad}}{\dot{p}}= H\left(\dfrac{\delta p}{\dot{p}} - \dfrac{\delta \varrho}{\dot{\varrho}}\right).
\end{equation}
Combining this definition for $\mathcal{S}$ with the continuity equation~\eqref{eq:hydro_continuity}, we can re-express Eq.~\eqref{R-dot-eq-full} as follows:
\begin{equation}
\label{eq:hydro_R-dot_S}
\dot{\cal R} = -3H\frac{\dot{p}}{\dot{\varrho}} \mathcal{S} + \dfrac{H}{\dot{H}}\dfrac{k^2}{a^2} \left( \frac{\dot{p}}{\dot{\varrho}}  \Psi + \frac12\big(\Psi - \Phi\big) \right).
\end{equation}
This relation together with Eq.~\eqref{eq:hydro_R_zeta_relation} show that $\mathcal{S}$ acts as a direct source of $\mathcal{R}$ and $\zeta$ on super-Hubble scales even when the gradient term is negligible ($k \rightarrow 0$), in which case
\begin{equation}
\label{master-eq}
\dot{\cal R} \approx -\dot{\zeta} \approx -3H\frac{\dot{p}}{\dot{\varrho}} \mathcal{S}.
\end{equation}

In some places, the adiabatic or density perturbations are characterized as `direct' sources of (co-moving) curvature modes ${\cal R}$ as observed in the CMB while ${\cal S}$ is often being referred to as an `indirect' source, a terminology that leads some to think that entropic sourcing is somehow more contrived.  
This semantic distinction is largely misleading, though, because its sole basis is the fact that, on super-Hubble scales, ${\cal R}$ and $\zeta$ are related to the (adiabatic) energy density fluctuation $\delta \varrho_{\rm N}$ by a zeroth-order relation~\eqref{eq:hydro_zeta} while they are related to ${\cal S}$ by a first-order relation~\eqref{master-eq}. In the proper physical sense, there is nothing indirect about sourcing curvature modes by the entropy fluctuation ${\cal S}$. 
Furthermore, as emphasized by Bardeen \cite{Bardeen:1980kt}, a generic fluid will have non-zero $\delta p_{\rm nad}$ and will include both adiabatic and entropic contributions to  ${\cal R}$.  The fact that the first examples studied in inflation had $\delta p_{\rm nad}=0$, is the oddity, as noted in \cite{Bardeen:1983qw}.   Furthermore, it is not necessary to have a fluid with multiple components to have $\delta p_{\rm nad}$ non-zero \cite{Bardeen:1980kt}.
For this reason, we will not employ the `direct-indirect' terminology.

\section{Curvature modes from `graceful exit'}
\label{sec:micro}

In bouncing and cyclic scenarios, where the universe is flattened and smoothed during a phase of slow contraction \cite{Khoury:2001wf}, a spectrum of quantum fluctuations in the stress-energy generated on sub-Hubble scales  ends up on super-Hubble scales by the end of slow contraction and classicalizes. This is due to the fact that the wavelength of these modes is proportional to $a(t)$, which shrinks at a much slower rate than the Hubble radius, as detailed in the Introduction. Notably, though, the amplitudes of adiabatic matter and metric fluctuations decay as contraction proceeds \cite{Creminelli:2004jg}.  

In this section, we demonstrate how non-adiabatic perturbations of the stress-energy generated during the slow contraction phase thereby become the natural source of co-moving curvature modes on super-Hubble scales during the stage of graceful exit, {\it i.e.}, when slow contraction ends and the transition to the bounce stage begins.

\subsection{Graceful exit in non-singular bouncing cosmologies}

A simple microphysical realization of slow contraction is a set up in which the stress-energy is supplied by two scalar fields, $\phi, \chi$ , minimally-coupled to Einstein gravity \cite{Li:2014qwa, Levy:2015awa},
\begin{equation}
\label{eq:background_action}
T_{\mu\nu}{}^S =  \nabla_\mu\phi \nabla_\nu\phi + \Sigma_1(\phi)\nabla_\mu\chi \nabla_\nu\chi
- \frac12 \Big(  \nabla_{\lambda} \phi \nabla^{\lambda} \phi +  \Sigma_1(\phi)\nabla_{\lambda}\chi \nabla^{\lambda}\chi
+ 2V(\phi, \chi) \Big) g_{\mu\nu}
\,.
\end{equation}
During the contracting phase $V \equiv V_0e^{-\phi/M}$ with $V_0<0$ and $\Sigma_1(\phi) \equiv e^{\phi/m}$ with $M$ and $m$ being the characteristic scales of the potential $V$ and the non-linear sigma type kinetic interaction $\Sigma_1$, respectively.

It is straightforward to show that $\nabla_\mu\chi \equiv 0$ is a solution of the corresponding system of Einstein-scalar field equations  and  that it is a stable fixed point solution \cite{Levy:2015awa}. This is because the kinetic interaction ($\propto \Sigma_1(\phi)$) modifies the second term in the $\chi$-equation,
\begin{equation}
\ddot{\chi} +  \left( 3H + \frac{\dot{\phi}}{m} \right)\dot{\chi}  = 0,
\end{equation}
flipping it from an anti-friction to a friction provided that $\dot{\phi}/m>|3H|$. As a result, the background evolution is driven solely by the $\phi$ field that acts like a perfect fluid on the background, rapidly and robustly homogenizing, isotropizing, and flattening the universe. As shown in Ref.~\cite{Ijjas:2020dws}, after only a few $e$-folds of contraction of $a(t)$, the dynamics is well-described by the FRW scaling attractor solution as specified in Eq.~\eqref{scaling-sol} with $M^{-1}=\sqrt{2\epsilon}$. 

Apart from perturbations in the $\chi$ field, all classical and quantum fluctuations of both the matter and the metric experience the Hubble anti-friction such that their amplitude is continuously decaying. This means, slow contraction is not just a classical but also a quantum smoother \cite{Cook:2020oaj}. At the same time, fluctuations in the $\chi$ field are being continuously damped and `see' a de Sitter like background. For the same reason as was argued in the inflationary context, the spectrum of perturbations in $\chi$ generated during slow contraction is nearly scale-invariant and  gaussian \cite{Levy:2015awa}. 

It is important to note, though, that perturbations in $\chi$ do {\it not} source co-moving curvature modes during the phase of slow contraction. The proof is based on evaluating the master equation~\eqref{eq:hydro_R-dot_S} derived above for the linearized Einstein-scalar system with the specific stress-energy $T_{\mu\nu}{}^S$ as given in Eq.~\eqref{eq:background_action}. 
With $\delta\phi \equiv \phi - \bar{\phi}$ and $\delta\chi  \equiv \chi - \bar{\chi}$ denoting the linearized scalar field components, the linearized energy density, pressure, and co-moving velocity perturbations take the following form:
\begin{eqnarray}
\label{eq:perturbs_p_rho_SF}
\delta \varrho_{\rm N} &=& -\delta T^0{}_0 = \dot{\phi}\delta\dot{\phi} 
+ \Sigma_1  \dot{\chi}\delta\dot{\chi}  
+ \Big({\textstyle \frac12} \Sigma_{1,\phi} \dot{\chi}^2 + V_{,\phi}\Big) \delta\phi 
-  \Big( \dot{\phi}^2 +\Sigma_1 \dot{\chi}^2\Big) \Phi
,\\
\label{eq:perturbs_p_SF}
\delta p_{\rm N} &=& \frac{1}{3}\delta T^i{}_i  =   \dot{\phi}\delta\dot{\phi} 
+ \Sigma_1  \dot{\chi}\delta\dot{\chi}  
+ \Big({\textstyle \frac12} \Sigma_{1,\phi} \dot{\chi}^2 - V_{,\phi}\Big) \delta\phi 
-  \Big( \dot{\phi}^2 +\Sigma_1 \dot{\chi}^2\Big) \Phi
,\\
\partial_i\delta u_{\rm N} &=& \big(p + \varrho\big)^{-1} \delta T^0{}_i = - \partial_i\left( \frac{\dot{\phi}\delta\phi + \Sigma_1 \dot{\chi}\delta\chi}{\dot{\phi}^2 + \Sigma_1 \dot{\chi}^2} \right) \, .
\end{eqnarray}
Here and onwards, we shall work in Newtonian gauge ($\sigma_{\rm sh} \equiv 0$) as the obvious gauge choice. For simplicity, we omit the bar for the background solutions $\phi(t)$ and $\chi(t)$.  
Using the Poisson-like equation~\eqref{eq:hydro_energy_momentum_constr} that we derived by combining the Hamiltonian and momentum constraints~(\ref{eq:hydro_Einst_00}-\ref{eq:hydro_Einst_0i}), we can identify the $k$-dependent components of the adiabatic pressure perturbation and obtain the simple expression
\begin{eqnarray}
\label{S-Sigma2=0}
\delta p_{\rm nad} &=&   \frac{ V_{, \phi} \dot{\phi} + V_{, \chi} \dot{\chi} }{3H\dot{H}} \left( -2 \frac{k^2}{a^2}\Psi \right)
+ 2 \Big(  V_{, \phi} \Sigma_1 \dot{\chi} -  V_{, \chi} \dot{\phi} \Big) \frac{\dot{\phi} \delta\chi - \dot{\chi} \delta \phi}{\dot{\phi}^2 + \Sigma_1 \dot{\chi}^2}
\,.
\end{eqnarray}
Manifestly, in the long-wavelength/super-Hubble limit ($k\ll aH$), there is no or no significant non-adiabatic pressure contribution if $V_{, \chi}\approx0$ and $\dot{\chi} \approx 0$, which is exactly the case for the fixed point attractor solution of interest.
It is for this reason that no co-moving curvature is being generated on super-Hubble scales during the smoothing phase, {\it i.e.}, with Eq.~\eqref{eq:hydro_R-dot_S}:
\begin{equation}
\label{rdot-s2=0}
\dot{\cal R} = \frac{H}{\dot{H}} \frac{k^2}{a^2}\Psi 
- \frac{H}{\dot{H}}  
\left(  V_{, \phi} \Sigma_1 \dot{\chi} -  V_{, \chi} \dot{\phi} \right) \times \frac{\dot{\phi} \delta\chi - \dot{\chi} \delta \phi}{\dot{\phi}^2 + \Sigma_1 \dot{\chi}^2}
\xrightarrow[k/aH \ll 1]{} 0
\,,
\end{equation}
which completes the proof that the $\chi$-perturbations cannot source curvature perturbations during slow contraction.

On the other hand, in what follows, we will show that the non-adiabatic super-Hubble fluctuations in the $\chi$ field generated during the phase of slow contraction {\it can} source co-moving curvature modes on super-Hubble scales {\it during graceful exit} from slow contraction. This fits naturally in scenarios involving a classical (non-singular) bounce. In these scenarios, the phase of slow contraction ends when the kinetic energy and the total energy density have both grown to be large enough such that the equation of state $\varepsilon$ shrinks and approaches 3, at which time higher-order kinetic terms in $\chi$ become important \cite{Ijjas:2016vtq,Ijjas:2019pyf}. We will demonstrate that such a set-up naturally leads to a non-trivial non-adiabatic pressure component that induces the generation of curvature modes on super-Hubble scales.

More specifically, to incorporate the stage of graceful exit leading up to the classical bounce phase, we extend the smoothing stress-energy component $T_{\mu\nu}{}^S$ in Eq.~\eqref{eq:background_action} as follows:
\begin{equation}
\label{eq:background_action-2}
T_{\mu\nu}{}^{\rm exit} = T_{\mu\nu}{}^S
+\frac14 \Sigma_2(\phi)\big(\nabla_{\lambda}\chi \nabla^{\lambda}\chi \big)^2 g_{\mu\nu}
- \Sigma_2(\phi)\big(\nabla_{\lambda}\chi \nabla^{\lambda}\chi \big)\nabla_{\mu}\chi \nabla_{\nu}\chi
,
\end{equation}
where $\Sigma_2 (\phi)$ is the quartic kinetic interaction. 

On an FRW background, the two fields {\it together} act as a perfect fluid with energy density and pressure taking the form
\begin{subequations}
\label{eq:background_p_rho}
\begin{alignat}{2}
\varrho  &= -T^0{}_0&&= \frac12 \dot{\phi}^2 + \frac12 \left(\Sigma_1 + \frac32 \Sigma_2\dot{\chi}^2\right)\dot{\chi}^2 + V
,\\
\label{eq:background_p_rho2}
p &= \frac{1}{3}T^i{}_i&&=\frac12 \dot{\phi}^2 + \frac12 \left(\Sigma_1 + \frac12 \Sigma_2\dot{\chi}^2\right)\dot{\chi}^2 - V;
\end{alignat}
\end{subequations}
such that the corresponding equation of state parameter is given by
\begin{equation}
\varepsilon = 3 \times \frac{\frac12 \dot{\phi}^2 + \frac12 \left(\Sigma_1 + \Sigma_2\dot{\chi}^2\right)\dot{\chi}^2}{\frac{1}{2}\dot{\phi}^2 
+ \frac12 \left(\Sigma_1 + \frac{3}{2}\Sigma_2\dot{\chi}^2\right)\dot{\chi}^2 + V} \,.
\end{equation}

The evolution of the background `fluid' is determined by the Euler-Lagrange equations for $\phi$ and $\chi$:
\begin{subequations}
\label{eq:background_eom_field}
\begin{align}
\label{full-phi-eq}
& \ddot{\phi} + 3H\dot{\phi}  + V_{,\phi} = \frac12 \left(\Sigma_{1,\phi} + \frac12 \Sigma_{2,\phi}\dot{\chi}^2\right) \dot{\chi}^2
,\\
\label{full-chi-eq}
& \left(\Sigma_1 + 3\Sigma_2\dot{\chi}^2\right)\ddot{\chi} + \left(\Sigma_1 + \Sigma_2\dot{\chi}^2\right)3H\dot{\chi} + \left(\Sigma_{1,\phi} + \Sigma_{2,\phi}\dot{\chi}^2\right)\dot{\phi}\dot{\chi} + V_{,\chi} = 0.
\end{align}
\end{subequations}
The quartic kinetic coupling in Eq.~\eqref{eq:background_action-2}is negligible during the smoothing phase when $\dot{\chi}\approx 0$ but becomes important as $a(t)$ continues to shrink and $\dot{\chi}^2$ increases. There, $\Sigma_2(\phi)$ plays a twofold role. First, by modifying the damping friction term in the $\chi$-equation~\eqref{full-chi-eq}, it drives the system away from the scaling attractor solution. In addition, the quartic term acts as an effective potential on the $\phi$-field in Eq~\eqref{full-phi-eq}, modifying its trajectory. Both effects together trigger the phase of graceful exit from the smoothing phase. 

\subsection{Sourcing curvature modes during graceful exit}

During graceful exit, the linearly perturbed energy density $\delta\varrho_{\rm N}$ and pressure $\delta p_{\rm N}$ components are given by 
\begin{subequations}
\label{eq:perturbs_p_rho}
\begin{alignat}{2}
\label{eq:perturbs_rho}
\delta \varrho_{\rm N} &= -\delta T^0{}_0 
&&= \dot{\phi}\delta\dot{\phi} + \Big(\Sigma_1 + 3\Sigma_2\dot{\chi}^2\Big)\dot{\chi}\delta\dot{\chi}
-  \Big(\dot{\phi}^2 + \left(\Sigma_1 + 3\Sigma_2\dot{\chi}^2\right)\dot{\chi}^2 \Big)\Phi 
\\
\nonumber& 
&&+\Big(V_{,\phi} + \frac{1}{2}\Sigma_{1,\phi}\dot{\chi}^2 + \frac34\Sigma_{2,\phi}\dot{\chi}^4\Big)\delta\phi + V_{,\chi}\delta\chi
,\\
\label{eq:perturbs_p}
\delta p_{\rm N} &= \frac{1}{3}\delta T^i{}_i &&=  \dot{\phi}\delta\dot{\phi} + \Big(\Sigma_1 + \Sigma_2\dot{\chi}^2\Big)\dot{\chi}\delta\dot{\chi}
-  \Big(\dot{\phi}^2 + \left(\Sigma_1 + \Sigma_2\dot{\chi}^2\right)\dot{\chi}^2 \Big)\Phi 
\\
\nonumber& 
&&-\Big(V_{,\phi} - \frac{1}{2}\Sigma_{1,\phi}\dot{\chi}^2 - \frac14 \Sigma_{2,\phi}\dot{\chi}^4\Big)\delta\phi - V_{,\chi}\delta\chi.
\end{alignat}
\end{subequations}
As before, the co-moving scalar velocity potential $\delta u_{\rm N}$ can be read off from the  $0i$-component of the linearized stress-energy tensor,
\begin{equation}
\delta T^0{}_i = -\dot{\phi}\partial_i\delta\phi - \left(\Sigma_1 + \Sigma_2\dot{\chi}^2\right)\dot{\chi}\partial_i\delta\chi = \left(p + \varrho\right)\partial_i\delta u_{\rm N},
\end{equation}
or equivalently,
\begin{equation}
\label{eq:perturbs_vel_potential}
\delta u_{\rm N} = -\dfrac{\dot{\phi}\delta\phi + \left(\Sigma_1 + \Sigma_2\dot{\chi}^2\right)\dot{\chi}\delta\chi}{\dot{\phi}^2 + \left(\Sigma_1 + \Sigma_2\dot{\chi}^2\right)\dot{\chi}^2}.
\end{equation}

To identify the $k$-dependent contributions to the non-adiabatic pressure component $\delta p_{\rm nad}$, we shall again employ the Poisson-like equation~\eqref{eq:hydro_energy_momentum_constr} which specified for $T_{\mu\nu}{}^{\rm exit}$ takes the form
\begin{eqnarray}
\label{eq:perturbs_energy_momentum_constr}
-2\,\dfrac{k^2}{a^2}\Psi & = & \dot{\phi}\delta\dot{\phi}  + \Big(\Sigma_1 + 3\Sigma_2\dot{\chi}^2\Big)\dot{\chi}\delta\dot{\chi} 
 - \Big(\dot{\phi}^2 + \Sigma_1 \dot{\chi}^2 + 3\Sigma_2\dot{\chi}^4\Big) \Phi 
\\
&+& \Big(\Sigma_{1,\phi} + \frac{3}{2}\Sigma_{2,\phi}\dot{\chi}^2\Big) \frac{1}{2}\dot{\chi}^2\delta\phi
+ V_{,\phi}\delta\phi + V_{,\chi}\delta\chi 
+3H\Big(\dot{\phi}\delta\phi + \big(\Sigma_1 + \Sigma_2\dot{\chi}^2\big)\dot{\chi}\delta\chi\Big)
\,.
\nonumber
\end{eqnarray}
Unlike scalar-field induced stress-energy forms that only involve quadratic kinetic terms, the right hand side of Eq.~\eqref{eq:perturbs_energy_momentum_constr} explicitly depends not only on the linearized scalar field components but also on the linearized lapse perturbation $\Phi$. We will use the momentum conservation equation~\eqref{eq:perturbs_conservation-M} to eliminate $\Phi$ and write it in terms of the matter field quantities $\delta \varrho_{\rm N}, p_{\rm N}$ and $\delta p_{\rm nad}$.

Substituting the expressions~(\ref{eq:perturbs_rho}-\ref{eq:perturbs_p}) for the linearized energy density $\delta \varrho_{\rm N}$ and pressure $\delta p_{\rm N}$ into the definition of $\delta p_{\rm nad}$ as introduced in Eq.~\eqref{eq:hydro_non-adiabatic_press} together with Eqs.~(\ref{eq:perturbs_conservation-M}, \ref{eq:perturbs_energy_momentum_constr}) yields
\begin{eqnarray}
\label{eq:perturbs_non-adiabatic_press}
\frac12 \left( 1 + \frac{ 2\Sigma_2\dot{\chi}^4}{\varrho + p}  \right) \delta p_{\rm nad} &= 
& \left( 1- \frac{\dot{p}}{\dot{\varrho}} \left( 1 + \frac{2 \Sigma_2\dot{\chi}^4}{\varrho + p}  \right) \right) \left( - \frac{k^2}{a^2}\Psi \right) 
\\
&+& \, \dfrac{ V_{,\chi}\dot{\phi} - \left(\Sigma_1 + \Sigma_2\dot{\chi}^2\right)\left(V_{,\phi} + \frac{1}{4}\Sigma_{2,\phi}\dot{\chi}^4\right)\dot{\chi} } {\varrho + p} \Big( \dot{\chi}\delta\phi - \dot{\phi}\delta\chi  \Big)
\nonumber\\
&-& \Sigma_2\dot{\chi}^3 \frac{d}{dt}\Big( \dot{\chi} \delta u_{\rm N} + \delta \chi  \Big)
\,,\nonumber
\end{eqnarray}
where $\varrho, p$ and $\delta u_{\rm N}$ are given by Eqs.~(\ref{eq:background_p_rho}-\ref{eq:background_p_rho2}, \ref{eq:perturbs_vel_potential}), respectively.
The first line of Eq.~\eqref{eq:perturbs_non-adiabatic_press} includes the gradient term. This is the only non-zero term in $\delta p_{\rm nad}$ and in $\mathcal{S} = H\,\delta p_{\rm nad}/\dot{p}$ in the case of a single scalar field; the fact that it is vanishing on large scales ($k/aH \ll 1$) is why $\mathcal{R}$ approaches a constant on super-Hubble scales in these models.
The second line originates from the variation of the potential $V$, and the quartic kinetic coupling function $\Sigma_2$. Finally, the third line follows from the perturbation of the quartic kinetic term, which, in contrast to the previous terms, also involves time derivatives of the scalar field perturbations. 

Finally, with Eq.~\eqref{eq:hydro_R-dot_S}, the time-variation of $\mathcal{R}$ can be re-expressed in the microscopic picture:
\begin{eqnarray}
\label{eq:perturbs_R-dot_fields}
\dot{\cal R} &=&
-  \frac{2H}{\dot{\phi}^2 + \Sigma_1\dot{\chi}^2 + 3\Sigma_2\dot{\chi}^4} \times \frac{k^2}{a^2}\Psi
\\
&+&  \frac{ \left(\Sigma_1 + \Sigma_2\dot{\chi}^2\right)\left( V_{,\phi} + \frac14\Sigma_{2,\phi}\dot{\chi}^4\right)\dot{\chi} - V_{,\chi}\dot{\phi} - \Sigma_2\dot{\chi}^3\ddot{\phi} }{\dot{\phi}^2 + \Sigma_1\dot{\chi}^2 + 3\Sigma_2\dot{\chi}^4} 
\times 2H  \frac{\dot{\phi}\delta\chi - \dot{\chi}\delta\phi}{\dot{\phi}^2 + \Sigma_1\dot{\chi}^2 + \Sigma_2\dot{\chi}^4}
\nonumber\\
&-&  \frac{ 2\Sigma_2\dot{\chi}^3\dot{\phi}}{\dot{\phi}^2 + \Sigma_1\dot{\chi}^2 + 3\Sigma_2\dot{\chi}^4}
\times H \frac{d}{dt} \left( \frac{\dot{\phi}\delta\chi - \dot{\chi}\delta\phi}{\dot{\phi}^2 + \Sigma_1\dot{\chi}^2 + \Sigma_2\dot{\chi}^4}\right)
\,,\nonumber
\end{eqnarray}
This is one of the main results of our paper. In contrast to the single-field case with conventional kinetic term, or the case of two scalar fields with quadratic kinetic coupling and $V=\Sigma_2=0$, the evolution of $\mathcal{R}$ is non-trivial on large scales in the presence of a second field together with a potential and a quartic kinetic term contributions.

\section{Geometric Interpretation (and the absence thereof)}

To illustrate the connection between the macroscopic (hydrodynamical) picture and the microphysics as well as the relation between the co-moving curvature and the entropy modes, ${\cal R}$ and ${\cal S}$, it is common to write ${\cal R}$ as a weighted sum of the individual contributions by the two perturbed scalars, 
\begin{equation}
\label{R-weighted}
{\cal R} \equiv \Psi - H \Big( \underbrace{ \omega_{\phi}\delta u_{\phi} + \omega_{\chi}\delta u_{\chi}}_{ \equiv \,\delta \sigma/\dot{\sigma}}\Big)  \,,
\end{equation}
which together define the adiabatic fluctuation $\delta \sigma$ with the adiabatic field defined through
\begin{equation}
\label{def-sigmadot} 
\dot{\sigma}\equiv \sqrt{-2\dot{H}}\,.
\end{equation}
Here $\delta u_{\phi}, \delta u_{\chi}$ are the co-moving scalar velocity potentials associated with the perturbed matter fields $\delta\phi$ and $\delta\chi$, respectively, given by
\begin{equation}
\delta u_{\phi} \equiv - \frac{\delta\phi}{\dot{\phi}},\quad \delta u_{\chi} \equiv - \frac{ \delta\chi}{\dot{\chi}}\,;
\end{equation}
and the weight factors are defined as
\begin{equation}
\omega_{\phi} \equiv \frac{\dot{\phi}^2}{-2\dot{H}}, \quad \omega_{\chi} \equiv \frac{\Sigma_1\dot{\chi}^2+\Sigma_2\dot{\chi}^4}{-2\dot{H}}\,.
\end{equation}
Note that $\omega_{\phi} + \omega_{\chi} \equiv 1$.

One can associate the weight factors with the angle $\vartheta$ between the scalars and the adiabatic direction in field space \cite{Gordon:2000hv}, {\it i.e.} 
 \begin{equation}
 \label{def-angles}
{\rm cos}^2 \vartheta \equiv \omega_{\phi}, \quad  {\rm sin}^2 \vartheta \equiv \omega_{\chi} \,.
\end{equation}
such that the adiabatic mode introduced in Eq.~\eqref{def-sigmadot} can be re-expressed as
\begin{equation}
\dot{\sigma} \equiv  {\rm cos}\vartheta \dot{\phi} + {\rm sin}\vartheta \dot{X} 
\end{equation}
and, similarly, the adiabatic fluctuation defined through Eq.~\eqref{R-weighted} can be re-written as
\begin{equation}
\delta \sigma = {\rm cos}\vartheta \, \delta\phi + {\rm sin}\vartheta  \,\delta X\,.
\end{equation}
Note that we rescaled $\dot{\chi}$ and $\delta \chi$ so that $\dot{X} \equiv \sqrt{\Sigma_1+\Sigma_2\dot{\chi}^2} \dot{\chi}$ and $\delta X \equiv \sqrt{\Sigma_1+\Sigma_2\dot{\chi}^2} \,\delta\chi
$.

In the case of two (or multiple) scalar fields with quadratic kinetic terms that have a well-defined field-space metric, this geometric interpretation suggests that the macroscopic entropy perturbation ${\cal S}$ as introduced in Eq.~\eqref{def-S-hydro} can be associated with a scalar field $\delta \sigma^{\perp}$ which, in field space, is perpendicular to the adiabatic direction $\delta \sigma$, {\it i.e.},
\begin{equation}
\label{delta-sigma-ast}
\delta \sigma^{\perp} \equiv -{\rm sin}\vartheta \, \delta\phi + {\rm cos}\vartheta  \,\delta X.
\end{equation}
For example, if $\Sigma_2\equiv0$ in Eq.~\eqref{eq:background_action-2} as is the case during slow contraction, the evolution equation for the adiabatic field $\sigma$ takes the simple form
\begin{equation}
\ddot{\sigma} + 3H \dot{\sigma} + V_{,\sigma}= 0\,,
\end{equation}
where the potential slope along the adiabatic field direction is given by
\begin{equation}
V_{,\sigma} \equiv {\rm cos}\vartheta \,V_{,\phi} + {\rm sin}\vartheta \, \frac{V_{,\chi}}{\sqrt{\Sigma_1}} .
\end{equation}
With Eq.~\eqref{def-angles}, it is immediately apparent that the non-adiabatic pressure perturbation $\delta p_{\rm nad}$ derived in Eq.~\eqref{S-Sigma2=0}  is proportional to $\delta \sigma^{\perp}$ in the long-wavelength limit ($k/aH \ll 1$),
\begin{equation}
\delta p_{\rm nad} \xrightarrow[k/aH\ll 1]{} - 2 \times \left(  {\rm cos}\vartheta \, \frac{V_{,\chi}}{\sqrt{\Sigma_1}} - {\rm sin}\vartheta \,V_{,\phi} \right) \times \delta \sigma^{\perp}\,,
\end{equation}
such that, on super-Hubble scales, the time variation of the co-moving curvature perturbation as obtained in Eq.~\eqref{rdot-s2=0} can be re-expressed as 
\begin{equation}
\dot{{\cal R}}  \xrightarrow[k/aH\ll1]{} 2 \times 
\left(  {\rm cos}\vartheta \, \frac{V_{,\chi}}{\sqrt{\Sigma_1}} - {\rm sin}\vartheta \,V_{,\phi} \right) \times \left( -H\frac{\delta \sigma^{\perp}}{\dot{\sigma}}\right) 
\,.
\end{equation}
Here, the proportionality term
\begin{equation}
 V_{,\sigma^{\perp}}\equiv {\rm cos}\vartheta \, \frac{V_{,\chi}}{\sqrt{\Sigma_1}} - {\rm sin}\vartheta \,V_{,\phi}
\end{equation}
can be viewed as the potential slope along the $\sigma^{\perp}$ direction.
That is, co-moving curvature modes are sourced by scalar {\it potentials} that have non-zero slopes along the $\sigma^{\perp}$ direction.
Since in the special case of two canonical scalars ($\Sigma_1\equiv1$), 
\begin{equation}
-\frac{V_{,\sigma^{\perp}}}{\dot{\sigma}} =  \dot{\vartheta}\,,
\end{equation}
this geometric picture led to relating the generation of non-adiabatic pressure with turning of the background trajectory in field space ($\dot{\vartheta}\neq0$). Indeed, cosmological model building in the context of slow contraction exclusively considered this option to source co-moving curvature modes by entropic fluctuations \cite{Gordon:2000hv,Lehners:2007ac}. This interpretation is misleading, though, because it only holds for  $\Sigma_1\equiv1$, as was noted, {\it e.g.}, in Ref.~\cite{DiMarco:2002eb}.

More generally, the geometric picture, according to which the entropic sourcing mechanism is always associated with a scalar potential that has a non-zero slope along the $\sigma^{\perp}$ direction, does {\it not} hold in scenarios where the transition to the expanding phase is driven by a classical (non-singular) bounce. As pointed out above, here slow contraction ends due to higher-order kinetic terms becoming important at high (yet classical) energy scales. 
 In these cases, $\dot{\mathcal{R}}$ depends on both $\delta \sigma^{\perp}$ and its time derivative $\delta \dot{\sigma}^{\perp}$ and can be sourced in the absence of potentials, as can be seen, {\it e.g.}, from Eq.~\eqref{eq:perturbs_R-dot_fields} and the examples presented below in Sec.~\ref{sec:special}.

This demonstrates an essential physical difference between the true \textit{hydrodynamical} entropy perturbation $\mathcal{S}$ and the \textit{microphysical} relative field fluctuation $\delta \sigma^{\perp}$, despite the nomenclature sometimes given to $\delta \sigma^{\perp}$ as the generalized entropy perturbation \cite{Gordon:2000hv,Wands:2002bn}.
To keep the physics clear, we suggest here and henceforth using `entropy perturbation' to refer to the unambiguous hydrodynamic quantity $\mathcal{S}$, and suggest using `relative field fluctuation' for the microphysical quantity $\delta \sigma^{\perp}$. 

\section{Special examples}
\label{sec:special}
In this section, we briefly discuss some of the ways curvature perturbation can be generated on super-Hubble scales using the entropic mechanism in cases where the scalar potential is negligible, $V(\phi, \chi)\approx 0$.

The scalar potential is the only source of coupling between the curvature perturbation $\mathcal{R}$ and the relative scalar matter field fluctuations  in the model of two fields with only quadratic kinetic terms ($\Sigma_2\equiv0$), leading to a non-zero non-adiabatic pressure contribution $\delta p_{\rm nad}\neq0$. The cases presented below where $V(\phi, \chi)$ may be negligibly small or zero illustrate the new possibilities arising due to the presence of higher-order kinetic terms as analyzed in this paper.  In particular, unlike the examples commonly presented in the literature, these scenarios do not rely on turning of the two-field background trajectory for the entropic mechanism to generate curvature perturbations.

If $V(\phi, \chi)\approx 0$, it follows from Eq.~\eqref{eq:perturbs_non-adiabatic_press} that the non-adiabatic pressure perturbation is given by
\begin{eqnarray}
\label{eq:perturbs_non-adiabatic_press-V=0}
\frac12 \left( 1 + \frac{ 2\Sigma_2\dot{\chi}^4}{\varrho + p}  \right) \delta p_{\rm nad} &\approx 
& \left( 1- \frac{\dot{p}}{\dot{\varrho}} \left( 1 + \frac{2 \Sigma_2\dot{\chi}^4}{\varrho + p}  \right) \right) \left( - \frac{k^2}{a^2}\Psi \right) 
\\
&+&  \left(  \frac14 \big(\Sigma_1 + \Sigma_2\dot{\chi}^2\big) \Sigma_{2,\phi}\dot{\chi}^5 - \Sigma_2\dot{\chi}^3\ddot{\phi} \right) \frac{\dot{\phi}\delta\chi - \dot{\chi}\delta\phi }{\varrho + p} 
\nonumber\\
&-& \Sigma_2\dot{\chi}^3 \frac{d}{dt}\Big(  \frac{\dot{\phi}\delta\chi - \dot{\chi}\delta\phi }{\varrho + p}  \Big)
\,,\nonumber
\end{eqnarray}
where the background energy density and pressure are defined as 
\begin{subequations}
\begin{alignat}{2}
\label{eq:background_p_rho}
\varrho  &\approx&& \frac12 \dot{\phi}^2 + \frac12 \Sigma_1\dot{\chi}^2 + \frac34 \Sigma_2\dot{\chi}^4
,\\
\label{eq:background_p_rho2}
p &\approx&& \frac12 \dot{\phi}^2 +\frac12 \Sigma_1\dot{\chi}^2 + \frac14 \Sigma_2\dot{\chi}^4 .
\end{alignat}
\end{subequations}
It is obvious from Eq.~\eqref{eq:perturbs_non-adiabatic_press-V=0} that, in general, $\delta p_{\rm nad}\neq0$ despite the absence of a potential interaction term between the scalars. As a matter of fact, non-adiabatic pressure is being sourced due to the higher-order kinetic term in $\chi$ even in the absence of kinetic coupling between the two scalars ({\it i.e.}, constant $\Sigma_1$ and $\Sigma_2$). 

For example, if all kinetic coefficients are constants, $\Sigma_1 \equiv 1$ and $\Sigma_2 \equiv \Lambda^{-4}$ (where $\Lambda$ is the strong coupling scale of the $\chi$-field), using Eqs.~(\ref{def-S-hydro}-\ref{eq:hydro_R-dot_S}), the time variation of the co-moving curvature fluctuation on super-Hubble wavelengths is given by
\begin{eqnarray}
\label{eq:perturbs_R-dot_fields_V=0,sigma12=const}
\dot{\cal R} \xrightarrow[k/aH\ll1]{} &-& \frac{2\Lambda^{-4}\dot{\chi}^3\ddot{\phi} }{\dot{\phi}^2 + \dot{\chi}^2 + 3\Lambda^{-4}\dot{\chi}^4} 
\times H  \frac{\dot{\phi}\delta\chi - \dot{\chi}\delta\phi}{\dot{\phi}^2 + \dot{\chi}^2 + \Lambda^{-4}\dot{\chi}^4}
\nonumber\\
&-&  \frac{ 2\Lambda^{-4}\dot{\chi}^3\dot{\phi}}{\dot{\phi}^2 + \dot{\chi}^2 + 3\Lambda^{-4}\dot{\chi}^4}
\times H \frac{d}{dt} \left( \frac{\dot{\phi}\delta\chi - \dot{\chi}\delta\phi}{\dot{\phi}^2 + \dot{\chi}^2 + \Lambda^{-4}\dot{\chi}^4}\right)
\,.\nonumber
\end{eqnarray}
The expression for $\dot{\cal R}$ in Eq.~\eqref{eq:perturbs_R-dot_fields_V=0,sigma12=const} illustrates the key novel features of the `kinetically driven' entropic mechanism: First, on large scales ($k/aH\ll1$), the co-moving curvature mode evolves as a function of both the relative field fluctuation and its time derivative. This is in sharp contrast to the cases in which $\mathcal{R}$ is sourced  due to the presence of a scalar interaction potential $V(\phi,\chi)$ between $\phi$ and $\chi$. Second, although the two fields do not interact directly, $\mathcal{R}$ is nevertheless being sourced on super-Hubble scales due to the presence of the quartic kinetic term.

The role of the non-negligible quartic kinetic coupling becomes further obvious when introducing a non-linear sigma-type interaction between the scalars ($\Sigma_1 \equiv \Sigma_1(\phi)$) but no quartic interaction  ($\Sigma_2 \equiv \Lambda^{-4}$). In this case,  the expression for $\dot{\mathcal{R}}$ remains the same as in the previous example where there was no direct interaction between the scalar matter fields:
\begin{eqnarray}
\label{eq:perturbs_R-dot_fields_V=0,sigma2=const}
\dot{\cal R} \xrightarrow[k/aH\ll1]{} &-&
  \frac{  2\Lambda^{-4}\dot{\chi}^3\ddot{\phi} }{\dot{\phi}^2 + \Sigma_1\dot{\chi}^2 + 3\Lambda^{-4}\dot{\chi}^4} 
\times H  \frac{\dot{\phi}\delta\chi - \dot{\chi}\delta\phi}{\dot{\phi}^2 + \Sigma_1\dot{\chi}^2 + \Lambda^{-4}\dot{\chi}^4}
\nonumber\\
&-&  \frac{ 2\Lambda^{-4}\dot{\chi}^3\dot{\phi}}{\dot{\phi}^2 + \Sigma_1\dot{\chi}^2 + 3\Lambda^{-4}\dot{\chi}^4}
\times H \frac{d}{dt} \left( \frac{\dot{\phi}\delta\chi - \dot{\chi}\delta\phi}{\dot{\phi}^2 + \Sigma_1\dot{\chi}^2 + \Lambda^{-4}\dot{\chi}^4}\right)
\,.\nonumber
\end{eqnarray}

On the other hand,  if the two scalars interact only through the quartic kinetic term ($\Sigma_1 \equiv 1$ and $\Sigma_2 \equiv \Sigma_2(\phi)$), the expression for the time variation of the co-moving curvature mode on super-Hubble scales is modified:
\begin{eqnarray}
\label{eq:perturbs_R-dot_fields_V=0,sigma1=const}
\dot{\cal R} \xrightarrow[k/aH\ll1]{} &&
  \frac{ \frac12\left(1 + \Sigma_2\dot{\chi}^2\right) \Sigma_{2,\phi}\dot{\chi}^5  - 2\Sigma_2\dot{\chi}^3\ddot{\phi} }{\dot{\phi}^2 + \dot{\chi}^2 + 3\Sigma_2\dot{\chi}^4} 
\times H  \frac{\dot{\phi}\delta\chi - \dot{\chi}\delta\phi}{\dot{\phi}^2 + \dot{\chi}^2 + \Sigma_2\dot{\chi}^4}
\nonumber\\
&-&  \frac{ 2\Sigma_2\dot{\chi}^3\dot{\phi}}{\dot{\phi}^2 + \dot{\chi}^2 + 3\Sigma_2\dot{\chi}^4}
\times H \frac{d}{dt} \left( \frac{\dot{\phi}\delta\chi - \dot{\chi}\delta\phi}{\dot{\phi}^2 + \dot{\chi}^2 + \Sigma_2\dot{\chi}^4}\right)
\,.\nonumber
\end{eqnarray}

\section{Summary and outlook}
\label{sec:concl}

Our goal in this paper was to explore the `entropic mechanism' for inducing the generation of co-moving curvature perturbations on large scales when the mode wavelengths are much larger than the Hubble radius ($k/aH\ll1$). The entropic mechanism is of particular importance in the context of bouncing and cyclic cosmologies in which the generation of a nearly scale-invariant spectrum of curvature perturbations during a slow contraction smoothing phase is not possible through a purely adiabatic mechanism.


The key lessons from this study are:
\begin{enumerate}
\item[-] by employing a hydrodynamic approach, we showed that co-moving curvature perturbations can be sourced whenever the non-adiabatic component of the linearized pressure is non-vanishing on super-Hubble scales; 
\item[-] this sourcing mechanism can be viewed as a consequence of local momentum conservation of the stress-energy in the context of Einstein gravity;
\item[-] two interacting scalar matter fields minimally coupled to Einstein gravity can act as a source for non-adiabatic pressure perturbation;
\item[-] a possible, already explored way for the generation of non-adiabatic pressure perturbations is the family of two-field models where the fields are coupled through their potential energy density \cite{Khoury:2001wf,Finelli:2002we,Lehners:2007ac,Buchbinder:2007ad,Koyama:2007ag};
\item[-] another, novel way analyzed in this paper is the `kinetic sourcing' where non-adiabatic pressure is generated due to the presence of higher-order kinetic terms;
\item[-] `kinetic sourcing' can occur even if the potential energy density of the fields is zero or negligible or if the fields do not interact through their kinetic terms ;
\item[-] the `kinetic sourcing' is especially important in the context of non-singular bouncing and cyclic cosmologies during the phase of graceful exit in cases where the smoothing stage ends due to higher-order kinetic terms becoming important.
\end{enumerate}

Our results open up new avenues for model building in various cosmological applications. In a forthcoming companion paper~\cite{Roman_paper2}, we  show under which conditions the novel `kinetically driven' entropic mechanism leads to the generation of a nearly scale-invariant spectrum of co-moving curvature perturbations consistent with current experimental observations. It would be especially interesting to see the implementation of the kinetic mechanism in other early-universe scenarios, such as models involving galilean genesis, ghost condensation or limiting curvature.

\section*{Acknowledgements}
We thank Paul J. Steinhardt and V. (Slava) Mukhanov for helpful comments and discussions. The work of A.I. is supported by the Lise Meitner Excellence Program of the Max Planck Society and by the Simons Foundation grant number 663083.
R.K. thanks the Max Planck Institute for Gravitational Physics (Hannover) for hospitality, where parts of this work were completed.

\bibliographystyle{plain}
\bibliography{bib_paper1}

\end{document}